\documentclass[preprint,showpacs,aps]{revtex4}
\usepackage{graphicx}
\begin{document}
\title{Free boundary problems describing two-dimensional pulse recycling and
motion in semiconductors}
\author{L. L. Bonilla \cite{LLB}}
\affiliation{Departamento de Matem\'aticas, Universidad Carlos
III de Madrid, Avda.\ Universidad 30, E-28911 Legan{\'e}s, Spain\\
Also: Unidad Asociada al Instituto de Ciencia de Materiales (CSIC),
28049 Cantoblanco, Spain}

\author{ R. Escobedo\cite{RE}}

\affiliation{MIRIAM (MIlan Research center for Industrial and Applied
Mathematics)\\
Dipartimento di Matematica, Universit\`a di Milano, Via
Saldini 50; 20133 Milano, Italy}

\author{ F. J. Higuera\cite{FJH} }

\affiliation{ETS Ingenieros Aeron\'auticos, Universidad Polit\'ecnica de
Madrid, Pza. Cardenal Cisneros 3, 28040 Madrid, Spain. }

\date{ \today  }

\begin{abstract}

An asymptotic analysis of the Gunn effect in two-dimensional samples of bulk
n-GaAs with circular contacts is presented. A moving pulse far from
contacts is approximated by a moving free boundary separating regions where
the electric potential solves a Laplace equation with subsidiary boundary
conditions. The dynamical condition for the motion of the free boundary is a
Hamilton-Jacobi equation. We obtain the exact solution of the free boundary
problem (FBP) in simple one-dimensional and axisymmetric geometries. The
solution of the FBP is obtained numerically in the general case and
compared with the numerical solution of the full system of equations. The
agreement is excellent so that the FBP can be adopted as the basis for an
asymptotic study of the multi-di\-men\-sional Gunn effect.

\end{abstract}

 \maketitle

\section{Introduction}
\label{sec:intro}
Excitable media exhibit a large response to a sufficiently strong disturbance
from their only stable stationary homogeneous state. This feature makes them
ideally suited to sustain propagation of pulses or wave trains \cite{mur93}.
Examples are the propagation of an action potential along the axon of a nerve
\cite{kee98}, the propagation of a grass fire on a prairie, pulse propagation
through cardiac cells \cite{kee98}, reaction-diffusion \cite{kur84} or
ecological systems \cite{mur93}. Semiconductor systems displaying negative
differential resistivity in their current-field characteristics are also
excitable systems albeit they have peculiar features due to the long range
character of the electromagnetic interaction \cite{nie95}. Thus {\em dc}
voltage bias conditions lead to pulse recycling (at contacts) and motion that
give rise to self-sustained oscillations of the electric current, the
so-called Gunn effect, \cite{gun63}. While most of the theoretical and
experimental studies of these phenomena deal with one dimensional geometries
of samples with attached planar contacts, recent experiments \cite{wil94} and
numerical studies \cite{bon01,tesis} have considered rectangular samples with
point contacts. In this case, many unusual oscillatory patterns are found
\cite{bon01,tesis}.

A large part of the literature on pulse propagation is devoted to
the mathematical description of their motion on one-dimensional
unbounded domains. In the case of self-oscillations in
semiconductor systems, such a description is the basis of
asymptotic analyses of pulse recycling and motion, both in
one-dimensional (1D) \cite{hig92,bon97} or axisymmetric
two-dimensional (2D) samples \cite{bon01,beh01}. These studies
exploit that the electric field has only one relevant component
whose integral yields the voltage difference between contacts.
The situation is very different in more general geometries and
new ideas need to be brought in. In this paper, we reduce pulse
propagation (far from contacts) to the motion of a free boundary
(FB) separating regions where the electric potential is a
harmonic function. The FB obeys a Hamilton-Jacobi equation (HJE).
On the FB, continuity and jump conditions hold, and additional
conditions on contacts and sample boundaries are needed for the
problem of finding the FB (free boundary problem or FBP) to have a
unique solution. On simple 1D and axisymmetric geometries, the
HJE can be solved exacly. Its solution describes very well the
motion of a discontinuity of the electric potential representing
the pulse far from the boundaries, as comparison with the
numerical solution of the full system of differential equations
shows. This is also true of the general 2D case, but now the
solution of the FBP has to be obtained numerically. In all cases,
recycling and annihilation of pulses at contacts have to be
described separately from the FB motion.

The rest of the paper is organized as follows. In Section \ref{sec:2}, we
present the governing equations of the Kroemer model for the Gunn effect in
two-dimensional samples of n-GaAs. An asymptotic derivation of the
FBP is given in Section \ref{sec:3}. Section \ref{sec:4} contains
the exact solutions of the FBP in the 1D and axisymmetric cases.
Numerical solutions of the FBP in the general 2D case and
comparisons with the numerical solution of the full system of
equations are presented in Section \ref{sec:5}. The last Section
contains our conclusions.

\section{Equations and boundary conditions}
\label{sec:2}
The Kroemer model \cite{kro66} consists of the following equations and
boundary conditions (in dimensionless units) for the concentration of free
carriers (electrons), $n$, and the electric potential, $\varphi$:
\begin{eqnarray}
\frac{\partial n}{\partial t} + \vec{\nabla} \cdot (n \vec{v}
- \delta \vec{\nabla} n) = 0, \label{1}\\
\vec{\nabla}^2 \varphi = n - 1, \label{2}\\
\vec{v}(\vec{E}) = \vec{E}\, \frac{1+ v_{s} E^{3}}{1+E^{4}}\,,
\label{3}\\
\vec{x} \in \Sigma_{c} : \enspace
\vec{E} \cdot \vec{N} = \rho\, (n \vec{v} - \delta \vec{\nabla} n) \cdot
\vec{N} \quad \mbox{ and } \quad \varphi = 0,
\label{4}\\
\vec{x} \in \Sigma_{a} : \enspace \vec{E} \cdot \vec{N} = \rho\, (n
\vec{v} - \delta \vec{\nabla} n) \cdot \vec{N} \quad \mbox{ and } \quad
\varphi = \Phi,  \label{5}\\
\vec{x} \in \Sigma_{o} : \enspace \vec{E} \cdot \vec{N} = 0 \quad \mbox{ and
} \quad  (n \vec{v} - \delta \vec{\nabla} n) \cdot \vec{N} = 0.
\label{outbdry}
\end{eqnarray}
Here (\ref{1}) and (\ref{2}) are the charge continuity and Poisson
equations, respectively. The dimensionless electric field is $\vec{E} =
\vec{\nabla} \varphi$ and $E = |\vec{E}|$. In these equations, the electron
density has been scaled with the uniform concentration of donor impurities
in the semiconductor, $N_D= 10^{15}$ cm$^{-3}$, and the electric field with
the field characterizing the intervalley transfer responsible for the
negative differential mobility involved in the Gunn oscillation, $E_R=3.1$
kV/cm. Distances and times have been measured with the dielectric length
and the dielectric relaxation time, $l_1 =\epsilon E_{R}/(e N_{D})
\approx 0.276\mu$m, $l_1/(\mu_0 E_R)\approx 1.02$ ps, respectively
($\mu_0$ is the zero-field electron mobility; see, {\it e.g.},
\cite{hig92} for details). The unit of electric potential is $E_R
l_1\approx 0.011$ V. The carrier drift velocity of Eq.\ (\ref{3}),
$\vec{v}(\vec{E})$, is already written in dimensionless units, and it has
been depicted in Fig.\ 1 of Ref. \onlinecite{bon01}. We assume that the
diffusion coefficient is constant, $\delta\approx 0.013$ (at 20K). In the
rest of the paper we assume also a zero saturation velocity: $v_s=0$. 

Boundary and bias conditions need to be imposed at the interfaces
between semiconductor and contacts, $\Sigma_{c,a}$, and on the
outer boundary of the semiconductor boundary $\Sigma_o$. Our
boundary conditions (\ref{4}) and (\ref{5}) assume that the
normal components of electron current density and electric field
are proportional at the semiconductor--contact boundary (Ohm's
law) \cite{hig92}, (in these equations, $\vec{N}$ is the unit
normal to $\Sigma_{c,a}$, directed towards the semiconductor).
For simplicity, we choose all contact resistivities $\rho$ to be
equal. Bias conditions are chosen to be $\varphi=0$ at the
cathode $\Sigma_{c}$ (injecting contact) and $\varphi= \Phi$ (the
applied voltage) at the anode $\Sigma_{a}$ (receiving contact).
If part of the semiconductor boundary does not have attached
contacts, the corresponding boundary conditions are zero flux
ones, as in Eq.\ (\ref{outbdry}). Typically $\delta >0$ is very
small, so that diffusion matters only inside boundary layers near
the contacts or inside thin shock waves~\cite{hig92,bon97}. The
latter are charge accumulations that will be treated simply as
discontinuities of the electric field~\cite{hig92}. Thus
diffusion effects may be left out of the conservation equation
(\ref{1}) when interpreting the results. If we set $\delta = 0$,
the first boundary condition in Eq.\ (\ref{5}) and the second one
in Eq.\ (\ref{outbdry}) should be omitted.

We can write an Amp\`ere's equation for the total current density
(electronic plus displacement), $\vec{j}$, by eliminating $n$ from (\ref{1})
using (\ref{2}):
\begin{eqnarray}
\vec{\nabla} \cdot \vec{j} = 0, \quad\quad\quad \mbox{ with } \nonumber\\
\vec{j} = (1+\vec{\nabla}^2 \varphi ) \vec{v} - \delta\, \vec{\nabla}
(\vec{\nabla}^2 \varphi) + \frac{\partial \vec{E}}{\partial t} \,.
\label{6}
\end{eqnarray}

\section{Derivation of the free boundary problem}
\label{sec:3}
Let us consider a rectangular sample with circular
contacts whose radii $r_c$ are large but much smaller than the
distance between contacts, $1\leq r_{c} \ll L$. The current
density varies slowly and follows adiabatically the electric
field profiles in the semiconductor except during brief periods in
which new pulses are shed from the cathodes. Close to a cathode
located at the origin, the electric field and the current density
are approximately axisymmetric and we can use the results of
Ref.~\onlinecite{beh01}. $\vec{j} = J \vec{r}/r^2$,
$r=|\vec{r}|$, $\vec{E} = E_1(J/r) \vec{r}/r$. $E_1(j)$ and
$E_2(j)$, with $E_1<E_2$, are the two positive zeros of the
function $v(E)-j$, with $v(E)= |\vec{v}(\vec{E})|$. The maximum
value of $|\vec{j}|$ during self-oscillations is somewhat larger
than $j_c = O(1)$ at which $E_2(j)=\rho j$. Correspondingly, the
maximum value of $J$ is $J_c=j_c r_c= O(r_c)$ and far from the
cathode, $r\gg r_c$, $J\ll r$ holds. This means that $E\sim
E_1(J/r) \approx J/r\ll 1$ and $v(E)\approx E$.
When $v_s=0$ the pulses move slowly over large regions of the
sample in which the field is stationary and small: $E\ll 1$.
Notice that $v(E)=E - E^5/ (1+E^4)$, which implies
$v(E)$ to be approximately linear on a wide range of field values,
$E^5\ll 1$. We conclude that $v(E)\approx E$ except near
the contacts and inside pulses. In these outer regions, space and
time derivatives can be neglected in Eq.\ (\ref{6}), which implies
$\vec{j} \approx \vec{v}(\vec{E})\approx \vec{E}$ there. Thus
div$\vec{j}=0$ yields $\vec{\nabla}^2 \varphi=0$ and the electric
potential $\varphi$ is a harmonic function outside pulses and
contact regions:
\begin{eqnarray}
\vec{\nabla}^2 \varphi = 0.   \label{7}
\end{eqnarray}

Let us now consider the pulse interior. A pulse is a narrow region of high
electric field bounded by a leading front and a trailing front which is a
shock wave. Outside the pulse $E\ll 1$ as explained before. The leading front
is a region at which $n=1+ \vec{\nabla}\cdot \vec{E} \approx 0$. Since we are
describing the pulse far from the contacts, $r\gg r_c\gg 1$, the electric
field is essentially normal to the pulse, $\vec{N}$. Then $E_N =\vec{E} \cdot
\vec{N} \approx r_w(t)-r$, where $r$ measures displacement along the normal
to the front and $r_w(t)$ yields the front location. The velocity of the
leading front is $dr_w/dt= j_N = \vec{j}\cdot \vec{N}$, according to Eq.\
(\ref{6}). The back of a triangular pulse of height $(E_+ - E_-)$ (the
trailing front) is a shock wave with speed given by the equal area rule
\cite{beh01}
\begin{eqnarray}
V(E_+,E_-)= {1 \over E_+ - E_-} \,\int_{E_-}^{E_+} v(E)\, dE 
\sim {\pi \over 4E_+},
\label{8}
\end{eqnarray}
where we have used that $E_- \sim E_1$ \cite{E1} as $E_+\gg 1$.
Then the trailing front velocity is small and small waves
move faster than large ones.
A key observation is that the pulse is {\em narrow} and it can be substituted
by a curve on a length scale of the order of the distance between contacts,
$L$. This is clear if leading and trailing fronts of the pulse are circular
\cite{beh01}. Then the bias $\Phi=O(L)$ is the integral of the electric field
from the cathode to the anode and the pulse width (equal to its height) is
$(E_+ - E_-) =O(\sqrt{\Phi})\ll \Phi$. In the general case, the pulses are
circular during a large part of their lives \cite{bon01} and we shall assume
that their widths remain much smaller than $L$ even when their shapes are no
longer circular. Then we assume that the pulses are curves $\Gamma$ given by
the equation:
\begin{eqnarray}
W(\vec{x},t)=0.
\label{10}
\end{eqnarray}

Clearly, there is a finite voltage drop across the pulse, $\sim E_+^2/2 =
O(\Phi)$, which means that the electric potential has a jump discontinuity at
$\Gamma$:
\begin{eqnarray}
{E_+^2\over 2} = [\varphi] \equiv \varphi_{+} - \varphi_{-}.
\label{11}
\end{eqnarray}
Here $\varphi_{-}$ and $\varphi_{+}$ are the limiting values of $\varphi$ as
$\vec{x}$ approaches $\Gamma$ from the region inside or outside $\Gamma$,
respectively. The relations div$\vec{j}=0$ and $\vec{j}\approx \vec{E}$ imply
that the normal component of the electric field (and therefore the normal
derivative of the electric potential) is continuous across $\Gamma$:
\begin{eqnarray}
j_N = (\vec{N}\cdot\vec{\nabla}\varphi)_{+} = (\vec{N} \cdot \vec{\nabla}
\varphi)_{-}. \label{12}
\end{eqnarray}

This $j_N$ is also the velocity of the leading front of the pulse along its
normal, which is nearly equal to that of the trailing front, $V$ given by
Eq.\ (\ref{8}), during most of the pulse lifetime. The pulse
velocity can also be obtained by differentiating Eq.\ (\ref{10}) with respect
to time:
$${\partial W\over \partial t} + \vec{\nabla}W\cdot {d\vec{x}\over dt} =0.$$
Since $\vec{N} = \vec{\nabla}W/|\vec{\nabla}W|$, the normal component of the
pulse velocity, $j_N$, is
\begin{eqnarray}
{d\vec{x}\over dt} \cdot \vec{N} = - {1 \over |\vec{\nabla} W|}\, {\partial W
\over \partial t }.   \label{13}
\end{eqnarray}

Using Eqs.\ (\ref{8}), $j_N=V$ and (\ref{13}), we obtain the
following equation for the position of the FB $\Gamma$:
\begin{eqnarray}
- {\partial W \over \partial t} =
 {\pi |\vec{\nabla} W| \over 4\sqrt{2[\varphi] \,}}
\quad \mbox{on} \quad W(\vec{x},t)=0.
\label{14}
\end{eqnarray}
Thus we have posed the following FBP:

{\it The electric potential $\varphi(\vec{x},t)$ is a harmonic function
inside and outside the FB $\Gamma$, with boundary conditions
(\ref{4}), (\ref{5}) and (\ref{outbdry}) on the semiconductor boundaries. On
the FB $\Gamma$, implicitly given by $W(\vec{x},t)=0$, $\varphi$
has a jump discontinuity $[\varphi]$ and its normal derivative satisfies
$$ (\vec{N}\cdot\vec{\nabla}\varphi)_{+} =
{\pi \over 4\sqrt{2 [\varphi]} }= (\vec{N} \cdot \vec{\nabla} \varphi)_{-} ,
$$
where $\vec{N} = \vec{\nabla}W /|\vec{\nabla}W|$. Furthermore, the
FB obeys the following HJE:
\begin{eqnarray*}
\qquad
- {\partial W \over \partial t} = 
{\pi |\vec{\nabla} W| \over 4\sqrt{2[\varphi] \,}}
\quad \mbox{on }\quad W=0.
\end{eqnarray*}

The conditions on the normal derivative of the electric potential at the
FB are equivalent to:
\begin{eqnarray*}
 (\vec{\nabla} \varphi \cdot \vec{\nabla} W)_+ =
{\pi |\vec{\nabla} W| \over 4\sqrt{2[\varphi] \,}}
= (\vec{\nabla} \varphi \cdot \vec{\nabla} W)_-
\end{eqnarray*}
on $W=0$.}

The HJE (\ref{14}) can be solved by the method of characteristics (the
Hamilton equations). To derive them, we just take a partial derivative of the
HJE with respect to $x$, and a partial derivative with respect to $y$. The
results are
\begin{eqnarray*}
{\partial \over \partial t}{\partial W \over \partial x} +
{\pi \over 4|\vec{\nabla} W| \sqrt{2[\varphi]} } \, \left( {\partial W
\over \partial x} {\partial^2 W\over \partial x^2} + {\partial W \over
\partial y} {\partial^2 W\over \partial x\partial y}\right)\\
= {\pi \over 8\sqrt{2 [\varphi]^3}} {\partial [\varphi]\over\partial x}
|\vec{\nabla} W| ,
\end{eqnarray*}
and a similar equation for $\partial W/\partial y$. The corresponding
characteristic equations for these first-order quasilinear partial
differential equations for $p= \partial W/\partial x$ and $q= \partial W/
\partial y$ are
\begin{eqnarray*}
{dx \over dt} & = & {{\pi \over 4\sqrt{2[\varphi]}} \over \sqrt{p^2 + q^2}}
\,  p,\\
{dy \over dt} & = & {{\pi \over 4\sqrt{2[\varphi]}} \over \sqrt{p^2 + q^2}}
\, q,\\
{dp\over dt} & = & - {\pi \over 8\sqrt{2[\varphi]^3}} {\partial [\varphi]\over
\partial s}\, q ,\\
{dq\over dt} & = & {\pi \over 8\sqrt{2[\varphi]^3}} {\partial [\varphi]\over
\partial s}\, p,\\
{dW\over dt} & = & 0.
\end{eqnarray*}
In these equations $s$ is arc length on the FB $\Gamma$, and we
have used that $\partial[\varphi]/ \partial x = - q (\partial[\varphi]/
\partial s)/\sqrt{p^2 + q^2}$ and $\partial[\varphi]/\partial y = p (\partial
[\varphi]/\partial s)/\sqrt{ p^2 + q^2}$ on $\Gamma$. These expressions can
be straightforwardly derived by using a local coordinate system on $\Gamma$
with basis vectors $\vec{N} = \vec{\nabla}W/|\vec{\nabla}W|$ and $\vec{T}=
(-\partial W/\partial y,\partial W/\partial x)/|\vec{\nabla}W|$. The jump
$[\varphi]$ depends only on the arc length on $\Gamma$ and $t$ because it is
defined only for $(x,y)\in \Gamma$ [these $(x,y)\in\Gamma$ have zero
projection onto $\vec{N}$]. The last equation for $W$ follows from the chain
rule, the Hamilton equations for $x$ and $y$ and the HJE:
\begin{eqnarray*}
{dW \over dt}= {\partial W \over \partial t} + {\partial W
\over \partial x} {dx\over dt} + {\partial W \over
\partial y} {dy\over dt} \\
= {\partial W \over \partial t} +
{\pi |\vec{\nabla}W| \over 4\sqrt{2[\varphi]}} = 0 .
\end{eqnarray*} 

The characteristic equations can be used to find $W(x,y,t)$ given an initial
condition $W(x_0,y_0,0)= W_0(x_0,y_0)$ such that the FB is
described initially by $W_0(x_0,y_0)=0$. Let us assume that $[\varphi]$ is a
known function of $s$ and $t$. In principle, we can find the solutions of the
above equations with initial data $x=x_0$, $y=y_0$, $p=\partial W_0/\partial
x_0$ and $q= \partial W_0/\partial y_0$. The result is a two-parameter family
of solutions $x= X(t;x_0,y_0)$, $y=Y(t;x_0,y_0)$. Let us assume that we can
invert this transformation for each $t>0$ (which should be true for $t$
sufficiently small), $x_0 = \xi(x,y,t)$, $y_0=\eta(x,y,t)$. The solution of
the HJE is $W(x,y,t)= W_0(\xi(x,y,t),\eta(x,y,t))$ because $W$ is constant
over the characteristics. Once $W$ is found for a given $[\varphi]$, the
Laplace equation can be solved for the electric potential in the different
regions of the sample separated by the FB $W=0$. Inserting these
solutions in the definition of the jump $[\varphi]$, we find an equation for
this jump. It seems clear that we can implement this procedure numerically to
device an explicit method such that $W$ and $[\varphi]$ are calculated at time
$t+\Delta t$ knowing their values at time $t$. In particular, we do not need
to find $(x_0,y_0)$ in terms of $(x,y,t)$. All we need is to know the
instantaneous location of the FB $W=0$, thus we only need to know
the evolution of those $(x_0,y_0)$ that are on $W_0=0$. For each $t>0$, the
locus of such $x= X(t;x_0,y_0)$, $y=Y(t;x_0,y_0)$ constitutes the
FB. More details on the numerical implementation of these ideas
are given in Section \ref{sec:5}.

The FBP describes the motion of a pulse far from contacts and other
boundaries or pulses. To obtain a complete asymptotic description of the Gunn
self-oscillations, we have to supplement its solution with a local
description of the field near the contacts and boundaries and a description
of pulse collisions. In particular, new pulses are shed from the cathodes as
the normal component of the current density there surpasses a critical value
$j_c$ which is the same as in the axisymmetric case \cite{beh01}. There are
cases in which two pulses collide and merge and cases in which a pulse
splits \cite{bon01}. In these cases our construction of the moving FB 
breaks down. What do we do then? Consider for instance two
circular pulses that become tangent at a point $(x_1,y_1)$ at
time $t_1>0$. Clearly there are two different initial points
$(x_0,y_0)$ that have evolved towards
$(x_1,y_1)$ and $W(x,y,t)$ is no longer univalued. Numerical simulations of
the complete system of equations show that the two pulses merge and adopt an
eight-shaped form; see Fig.\ 7 of Ref.~\onlinecite{bon01}. To mimic this
situation with our FBP, we should stop the simulations and start with a new
FB at $t=t_1$ that is an eight-shaped simple curve with a hole at
the tangent point of the two old pulses. The new FBP should now have a unique
solution.

\section{Exact solutions of the free boundary problem in simple geometries}
\label{sec:4}
There are two simple geometries in which the FBP can be solved exactly:
parallel planar contacts attached at the ends of a rectangular sample (1D
case) and the Corbino geometry of two concentric circular contacts with the
sample in between (axisymmetric case). Let us call region A that comprising
the cathode and region B that comprising the anode.

\subsection{1D geometry}
Then the electric potential depends only on the coordinate $x$, the cathode
is located at $x=0$, the anode at $x=L$ and the FB is a moving
point $x_s(t)$. The electric potential obeys
\begin{eqnarray*}
{\partial^2\varphi_A\over\partial x^2}=0 \,\, \mbox{in } (0,x_s),\,\,
  \varphi_A(0,t)=0,
{\partial\varphi_A\over\partial x}(x_s,t)={\pi \over 4 \sqrt{2
[\varphi]}} ;\\
 {\partial^2\varphi_B\over\partial x^2}=0 \,\,\mbox{in }
(x_s,L), \,\,  \varphi_B(L,t)=\Phi ,
{\partial\varphi_B\over\partial x}(x_s,t)={\pi \over 4 \sqrt{2 [\varphi]\,}}.
\end{eqnarray*}
The solutions are
\begin{eqnarray*}
\varphi_A(x,t) &=& {\pi \over 4 \sqrt{2[\varphi]\,}}\, x, \\
\varphi_B(x,t) &=& {\pi \over 4 \sqrt{2[\varphi]\,}}\, (x-L)+\Phi.
\end{eqnarray*}
The jump in the potential, $[\varphi]= \varphi_B(x_s,t) -\varphi_A(x_s,t)$ is
independent of $t$ and $x_s$, and it solves the following equation:
\begin{eqnarray*}
[\varphi]= \Phi - {\pi \over 4 \sqrt{2[\varphi]}}\, L.
\end{eqnarray*}
Setting $\alpha = \sqrt{[\varphi]}$ and $\phi = \Phi/L$, we obtain
\begin{eqnarray}
\alpha^3 = \left(\phi \, \alpha - {\pi \over 4 \sqrt{2}}\right) L. \label{16}
\end{eqnarray}
Depending on the values of $\phi$ and $L$ this equation may have zero, one or
two positive solutions. If there are two solutions, an argument due to Volkov
and Kogan \cite{vol69} shows that the pulse with smaller $[\varphi]$ is
unstable. The FB $x_s(t)$ can be found by solving the dynamical
HJE:
\begin{eqnarray*}
-{\partial W \over \partial t} = {\pi \over 4 \sqrt{2[\varphi]\,}}
\left|{\partial W \over \partial x} \right| .
\end{eqnarray*}
Let us assume that the initial profile $W(x,0)= W_0(x)$ is monotone
increasing and that it vanishes at a position $x_s(0)\in (0,L)$ corresponding
to the pulse location at time $t=0$. For small enough $t$, we then have
$\partial W/\partial x >0$ and we can ignore the absolute value in the
previous equation. Its solution is then
$$ W(x,t) = W_0\left(x - {\pi t\over 4 \sqrt{2[\varphi]\,}}\right).$$
Notice that we have $\partial W/\partial x >0$ for all $t>0$. Since
$W(x_s,t)=0$, the previous solution yields
\begin{eqnarray}
x_s(t) =  x_s(0) + {\pi \over 4 \sqrt{2 [\varphi] \,}}\, t.  \label{17}
\end{eqnarray}

\begin{figure}
\begin{center}
\includegraphics[width=7.5cm]{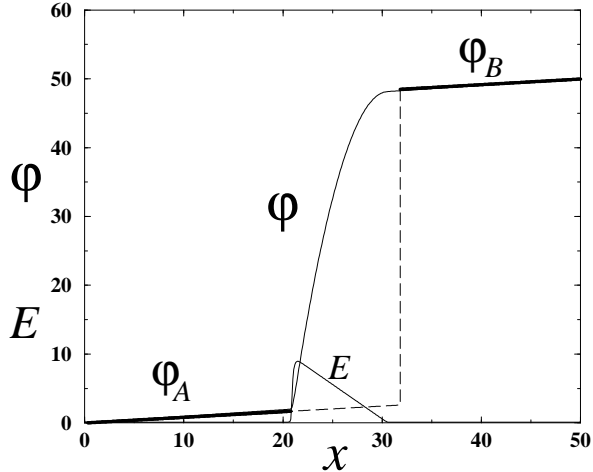}
\caption{The solid lines indicate the electric potential and field of an
advancing 1D pulse (far from the contacts) calculated by numerically
solving the Kroemer model. They agree very well with the approximations
$\varphi_A(x,t)$ and $\varphi_B(x,t)$ (dashed lines). We have used the
nondimensional units defined in the text.}
\label{fig1}
\end{center}
\end{figure}

Fig. \ref{fig1} compares $\varphi_A(x,t)$ and $\varphi_B(x,t)$ to the electric
potential of an advancing pulse calculated by numerically solving the exact
system of equations.

The FBP has yielded the same approximation to the complete
1D problem as indicated in Ref.~\onlinecite{hig92} for the motion of a pulse
far from the boundaries. When the pulse arrives at the anode $x=L$, it starts
disappearing there and the current density increases until it surpasses $j_c$.
Then a new pulse is shed at $x=0$; see Ref.~\onlinecite{hig92} for details.

\subsection{Corbino geometry (axisymmetric case)}
The potential depends only on the radius $r$ measured from the center of the
cathode. Solving the Laplace equation $\partial [r\, \partial \varphi/
\partial r]/ \partial r = 0$ at both sides of the moving pulse of radius
$r_s(t)$, we find
\begin{eqnarray*}
\varphi_A(r,t) & = &  {\pi \, r_s \over 4\sqrt{2[\varphi]\,}} \,
\log\left({r\over r_c}\right),\\
\varphi_B(r,t) & = & {\pi \, r_s \over 4\sqrt{2[\varphi]\,}} \,
\log\left({r\over r_c+L}\right) + \Phi.
\end{eqnarray*}
The jump in the electric potential at $r_s$ is now given by the following
equation:
\begin{eqnarray*}
[\varphi] = \Phi - {\pi \, r_s \over 4 \sqrt{2 [\varphi] \,}} \,
\log\left({r_c+L\over r_c}\right) ,
\end{eqnarray*}
or equivalently
\begin{eqnarray}
\alpha^3 =  \Phi \, \alpha - {\pi \over 4 \sqrt{2}} \, \log\left({r_c+L\over
r_c}\right)\, r_s. \label{18}
\end{eqnarray}
for $\alpha=\sqrt{[\varphi]}$. Notice that $r_s$ explicitly appears in these
equations and that $[\varphi]$ decreases as the pulse advances (and therefore
$r_s$ increases); cf.\ Ref.~\onlinecite{beh01}. The HJE can be solved as in
the 1D case and its solution yields
\begin{eqnarray}
r_s(t) =  r_s(0) + {\pi \over 4 \sqrt{2}}\,\int_0^t  [\varphi]^{-
{1\over 2}} dt . \label{19}
\end{eqnarray}
In this case, Eqs.\ (\ref{18}) and (\ref{19}) for $ [\varphi]$ and $r_s(t)$
need to be solved simultaneously.

The stage of a self-oscillation described by the previous FBP corresponds to
having a single pulse far from the contacts. See Ref.~\onlinecite{beh01} for
a fuller description of self-oscillations in this case.

\section{Numerical results}
\label{sec:5}
\begin{figure}
\begin{center}
\includegraphics[width=8.7cm]{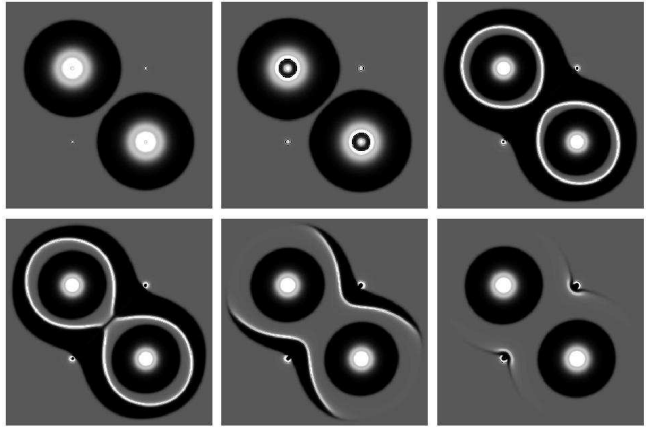}
\caption{Density plots of the solution of the Kroemer's model (with
$v_s=0$) in a square of side $l$=20 with four circular contacts forming
the vertices of a square of side $d=4$ located at the center of the sample.
Cathodes have potential $\varphi=0$ and anodes have $\varphi=10$.
Our dimensionless units have been defined in Section~\ref{sec:2}.}
\label{fig2}
\end{center}
\end{figure}

To test our FBP formulation, we shall consider the relatively complicated
geometry of Fig.~7 in Ref.~\onlinecite{bon01} (reproduced here as Fig.
\ref{fig2} to facilitate comparison with the results of numerically solving
the FBP) corresponding to $v_s=0$. The sample is a square of side $l=20$
with two cathodes at potential $\varphi=0$ and two anodes with $\varphi=
10$. The circular contacts (of radii 0.5) are at the vertices of a square
of side $d=4$ located at the center of the sample. Then the separation
between contacts is $L=3$ and the distance from contacts to the border of
the sample is 7.5. Notice that dipole waves are emitted from the cathodes.
Immediately after their emission, the waves are circular. As they approach
each other, the waves become elongated and merge forming an eight-shaped
connected curve that grows until it reaches the anodes.

\subsection{Free boundary problem}
\begin{figure}
\begin{center}
\includegraphics[width=8.7cm]{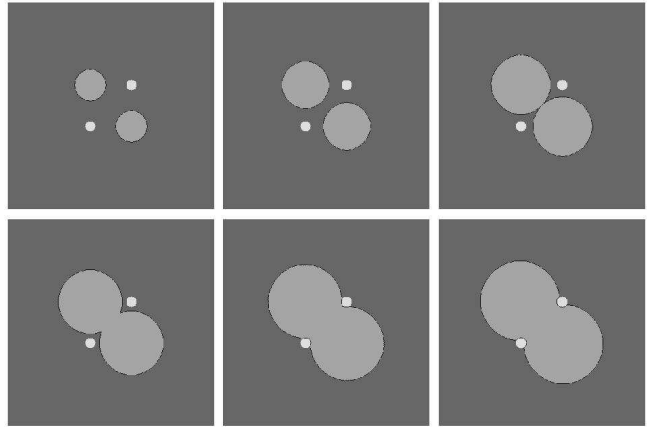}
\caption{Time evolution of the FB (black curve)
separating the two regions of the sample, inside (clear grey)
and outside (dark grey) the boundary. The anodes appear in white.}
\label{fig3}
\end{center}
\end{figure}

\begin{figure}
\begin{center}
\includegraphics[width=10cm]{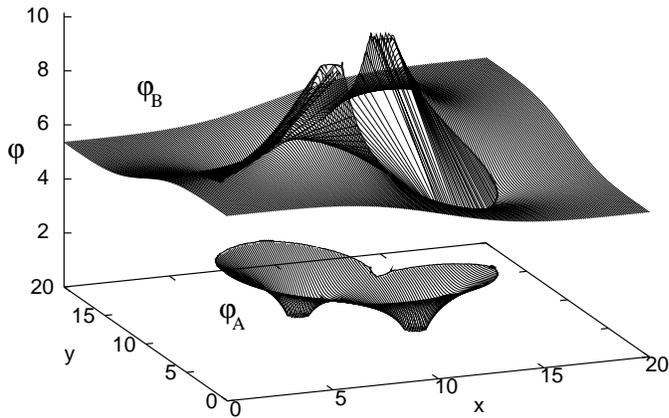}
\caption{3D plot of the electric potential surfaces $\varphi_A(x,y,t)$
(lower surface, inside the FB) and $\varphi_B(x,y,t)$ (upper
surface, outside the FB) at the time corresponding to the last
frame of Fig. \ref{fig3}.
Our dimensionless units have been defined in Section~\ref{sec:2}.}
\label{fig4}
\end{center}
\end{figure}

We shall now explain the results obtained by solving numerically the FBP.
Details of the numerical method will be given later. Fig. \ref{fig3} shows
the evolution of the FB separating the two regions of the sample, inside
and outside the boundary. Notice that the numerical solution of the FBP
closely resembles the numerical solution of the full Kroemer model depicted
in Fig.~\ref{fig2}. In the two first frames of Fig. \ref{fig3}, the FB
consists of two circumferences corresponding to the dipole waves nucleated
at the cathodes. In the third frame, the curves collide and then merge
forming an eight-shaped closed curve as shown in the remaining frames of
Fig. \ref{fig3}. Fig. \ref{fig4}  shows the electric potential distribution
in each region (inside and outside the FB) corresponding to the last frame
of Fig. \ref{fig3}.

\begin{figure}
\begin{center}
\includegraphics[width=9.5cm]{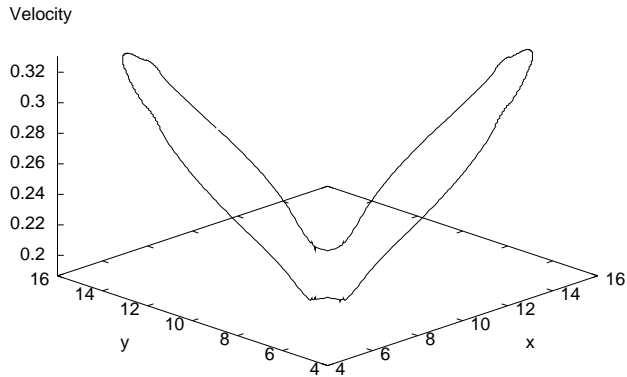}
\caption{Dimensionless velocity of each point of the FB at the dimensionless
time corresponding to the last frame in Fig. \ref{fig3} and calculated from
the electric potential distribution showed in Fig. \ref{fig4}.}
\label{fig5}
\end{center}
\end{figure}

By using Eq.\ (\ref{8}) we see that each point of the FB moves with
velocity
\begin{eqnarray*}
V = {\pi \over 4 \sqrt{2[\varphi]\,}}.
\end{eqnarray*}
Fig. \ref{fig5} depicts the velocity of the points at the FB in
the last frame of Fig. \ref{fig3}. The curve is symmetric and Fig.
\ref{fig5} shows that the FB moves faster at the points located
in the left-upper and right-lower corners of the sample, in agreement with
the numerical solution of the Kroemer model.

\begin{figure}
\begin{center}
\includegraphics[width=9.5cm]{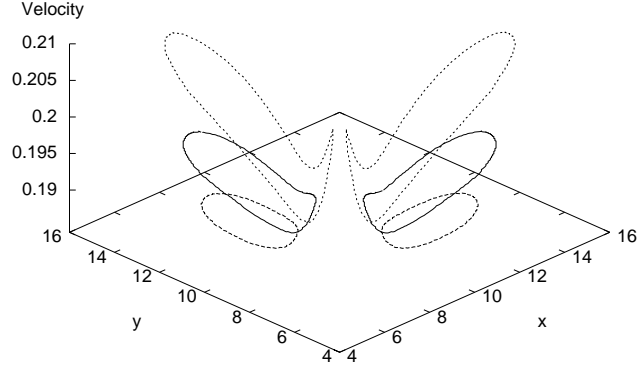}
\caption{Time evolution (from bottom to top) of the velocity
of the FB $\Gamma$ when $t<t_1$, where the topology is
composed by three domains and $\Gamma$ is made of two circumferences.
Our dimensionless units have been defined in Section~\ref{sec:2}.}
\label{fig6}
\end{center}
\end{figure}

\begin{figure}
\begin{center}
\includegraphics[width=9.5cm]{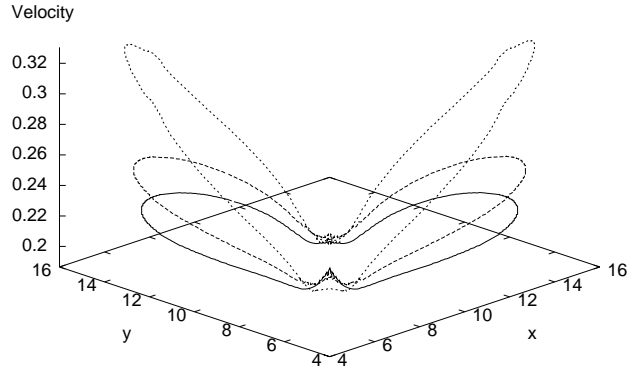}
\caption{Time evolution (from bottom to top) of the velocity
of the FB $\Gamma$ when $t>t_1$. The topology is now
composed by two domains.}
\label{fig7}
\end{center}
\end{figure}

Let $t_1$ be the time at which two dipole waves created at the cathodes
touch at a point (as in the third frame of Fig.\ \ref{fig3}), counted from
the time at which dipole waves are emitted at the cathodes ($t=0$). The
velocity of the points at the FB is shown at three different
times in Figures \ref{fig6} ($0<t<t_1$) and \ref{fig7} ($t>t_1$). Notice
that the velocity of the points near the center of the sample in Fig.
\ref{fig6} is larger than in neighboring points, which explains the
elongated form of the dipole waves in the numerical solution of the Kroemer
model (see  the third and fourth images of Fig. \ref{fig2}). In Fig.\
\ref{fig7} we observe that the largest velocity is reached at the outer
points of the single FB, also in agreement with the numerical
solution of the full model equations.

\subsection{Numerical solution of the free boundary problem}
To solve numerically the FBP, we should solve the PDE governing
the time evolution of the FB, taking into account that the
velocity thereof is determined by the solution of Laplace's
equation with Neumann boundary conditions on the FB and Dirichlet
boundary conditions at the contacts (the electric potential
problem, or, briefly, the EPP).

At each time step, the FB advances at a constant velocity for a
short distance from its previous position. (Thus we ignore the
velocity variation during the short time interval between $t_i$
and $t_i + \Delta t$). At time $t_i + \Delta t$, we solve the EPP
in the different domains resulting from the new location of the
FB. This yields the electric potential distribution that is used
to calculate the velocity of the FB at the next time step.

The time evolution of the free boundary is calculated by using
the so-called fast marching method (a special case of the method
of level sets). This method was introduced by Sethian in 1996
\cite{seth1} and used in a wide variety of applications
\cite{seth2,chopp,seth3}. Level sets methods are very efficient
for solving complex problems of evolving interfaces whose
topology may change. If the velocity of the interface does not
change sign, the fast marching method is a very fast algorithm
indeed. 

The general version of the method of level sets consists of
solving the evolution equation
\begin{eqnarray}
{\partial W \over \partial t} + F| \vec{\nabla} W| = 0,
\label{levelset-eq}
\end{eqnarray}
where $W(\vec{x},t)$ is a function such that $W$=$0$ describes the
free boundary moving at velocity $F$; cf.\ Eq.\ (\ref{14}). When
the sign of $F$ does not change, the FB either expands or
contracts uniformly as time elapses. In our case, the FB moves
away from the cathodes. Then the zero-level set $W$=$0$ comprises
the points farthest from the cathodes that have been traversed
once by the FB at a given instant of time. Then we can define an
{\it arrival time} function $T$ in the whole sample: $T(\vec{x})$
is the time it takes the FB to arrive at the point $\vec{x}$
starting from a given initial configuration. To find an equation
for $T$, we take the gradient of $W(\vec{x},T(\vec{x}))=0$, 
$\vec{\nabla} W + W_t \, \vec{\nabla} T = 0$, and use Eq.\
(\ref{levelset-eq}) to obtain 
\begin{eqnarray}
\vec{\nabla} W - F| \vec{\nabla} W| \, \vec{\nabla} T = 0.
\end{eqnarray}
This equation implies that $\vec{\nabla} W$ and $\vec{\nabla} T$
are colinear vectors and their lengths are related by $
|\vec{\nabla} W| = F| \vec{\nabla} W| \, |\vec{\nabla}T|$. 
Then we obtain the following Eikonal equation for $T$, 
\begin{eqnarray}
|\vec{\nabla} T(\vec{x})| = {1 \over F(\vec{x})}\equiv {4\sqrt{2\,
 [\varphi]}\over \pi}. \label{eikonal}
\end{eqnarray}
The velocity $F$ as a function of $\vec{x}$ is evaluated at time $t$.
Once the solution of Eq.~(\ref{eikonal}) is known at a narrow band
about the instantaneous location of the FB at time $t$, the location
thereof at time $t+ \Delta t$ is found by solving $T(\vec{x})=t+\Delta t$.

The fast marching method consists of solving numerically this
equation by using upwind finite differences to approximate
$|\vec{\nabla} T|$. In particular, we have used the Godunov scheme
\begin{eqnarray}
\max{ \left( {T_{i,j}-T_{i-1,j} \over \Delta x},{T_{i,j}-T_{i+1,j} \over
\Delta x},0 \right) }^2 + \nonumber \\
 \max{ \left( {T_{i,j}-T_{i,j-1} \over \Delta
y},{T_{i,j}-T_{i,j+1} \over \Delta y},0 \right) }^2 = {1 \over F_{i,j}^2}.
\end{eqnarray}
This choice ensures that the information is always taken from
where the solution is already known. The fast marching method
is consistent with the Huygens principle even when two waves
collide and adopt an eight-shaped curve as in Fig.~\ref{fig2}, or
with even more complex topologies. The EPP is solved by using an
integral equation method based upon Green's formula. This yields
the solution $\varphi$ within a region for a given value of its
normal derivative at each point of the boundary. To make sure
that the nonlinear boundary conditions at the FB hold, we
implement an iterative process.

\begin{figure}
\begin{center}
\includegraphics[width=8cm]{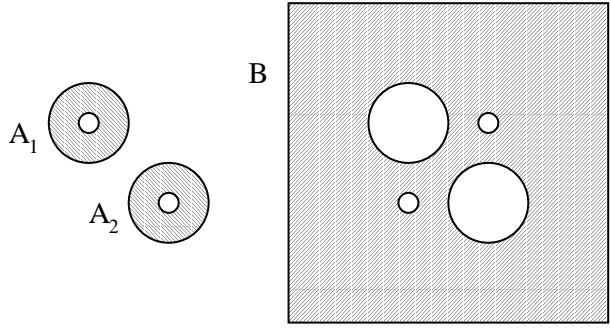}
\caption{The FB comprises two separate curves defining three
regions.}
\label{fig8}
\end{center}
\end{figure}

\begin{figure}
\begin{center}
\includegraphics[width=8cm]{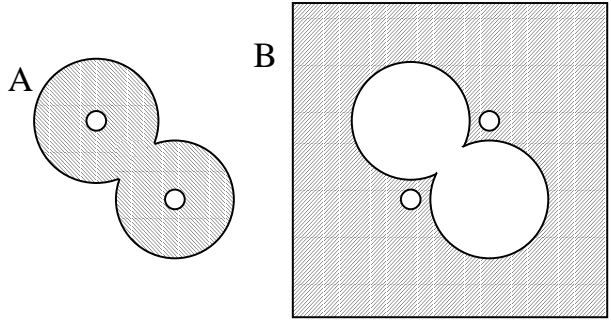}
\caption{The FB is a single curve defining two regions.}
\label{fig9}
\end{center}
\end{figure}

We shall start our simulation from an initial configuration as
depicted in Fig. \ref{fig8}. There two waves have been nucleated
at the cathodes and have reached their typical circular form. The
FB consists of two circumferences that divide the sample in three
regions, $A_1$, $A_2$ and $B$, in which we should simultaneously
solve the EPP. Implementing the fast marching method, we see two
waves growing from the initial circumferences until a time $t_1$,
when they meet at the center of the sample. Then the FB is a
connected curve and we have the situation depicted in Fig.
\ref{fig9}, where there are only two regions $A$ and $B$. The
algorithm detects the time $t_1$, adapts itself immediately to
the new configuration similar to Fig. \ref{fig9} and it
continues solving the FBP. 

The accuracy and convergence of the method have been successfully
checked by decreasing the mesh size. The computational cost of
the method is very low compared to the computational and memory
effort required by the resolution of the full Kroemer model. The
order-one fast marching method solves the Eikonal equation in the
whole sample with $O(N \log N)$ operations, where $N$ is the size
of the mesh, but we only need to solve the Eikonal equation in a
narrow band ahead of the FB at each time step. On the other hand,
the EPP solver carries out $O(N^2+M)$ operations, where $M$ is
the number of points defining the FB (at most of order $N$). 

\section{Conclusions}
\label{sec:6}
We have studied Gunn oscillations in 2D rectangular samples of
n-GaAs with circular contacts by solving the Kroemer
drift-diffusion model with appropriate boundary and initial
conditions. By using singular perturbation methods, the motion of
dipole waves in semiconductor samples has been reduced to solving
a free boundary problem. Exact solutions of this problem have
been found in simple 1D and axisymmetrical (Corbino) geometries.
In the general case, the free boundary is numerically found by
means of the fast marching method which is a special case of the
method of level sets. The great reduction in computational cost
allowed by using this method as an alternative to solving the
full Kroemer model would enable us to study much larger samples.

\acknowledgments
This work has been supported by the Spanish MCyT through grant
BFM2002-04127-C02-01, by the Third Regional Research Program of the
Autonomous Region of Madrid (Strategic Groups Action), and by the European
Union under grant HPRN-CT-2002-00282. We thank G.~Oleaga for fruitful
discussions.


\begin{thebibliography}{}
\bibitem{E1}
Outside the pulse, the field is quasistationary, so that
$E_-\approx E_1(j_N)$.
\end{thebibliography}
\end{document}